\documentclass[12pt, a4paper, reprint, aps, prd, nofootinbib, nobibnotes, floatfix, showkeys]{revtex4-1}

\usepackage[utf8]{inputenc}
\usepackage{amsmath,amssymb,amsthm,amsfonts}

\usepackage{graphicx}

\usepackage{mathtools}
\usepackage{physics}

\usepackage[colorlinks]{hyperref}

\newcommand{\orcidlink}[1]{}

\usepackage{booktabs}
\usepackage{parskip}
\usepackage{siunitx}

\newcommand{\james}[1]{}

\AtBeginDocument{
  \heavyrulewidth=.08em
  \lightrulewidth=.05em
  \cmidrulewidth=.03em
  \belowrulesep=.65ex
  \belowbottomsep=0pt
  \aboverulesep=.4ex
  \abovetopsep=0pt
  \cmidrulesep=\doublerulesep
  \cmidrulekern=.5em
  \defaultaddspace=.5em
}

\begin{document}

\title{Nondegenerate internal squeezing: an all-optical, loss-resistant quantum technique for gravitational-wave detection}

\author{James~W.~Gardner\,\orcidlink{0000-0002-8592-1452}}
\thanks{Corresponding author: \href{mailto:james.gardner@anu.edu.au}{james.gardner@anu.edu.au}.}
\author{Min~Jet~Yap\,\orcidlink{0000-0002-6492-9156}}
\author{Vaishali~Adya\,\orcidlink{0000-0003-4955-6280}}
\thanks{Now at the Department of Applied Physics, KTH Royal Institute of Technology, Roslagstullsbacken 21, Stockholm SE-106 91, Sweden.}
\author{Sheon~Chua\,\orcidlink{0000-0001-8026-7597}}
\author{Bram~J.~J.~Slagmolen\,\orcidlink{0000-0002-2471-3828}}
\author{David~E.~McClelland\,\orcidlink{0000-0001-6210-5842}}
\affiliation{
Centre for Gravitational Astrophysics, The Australian National University, Acton, ACT, 2601, Australia
}
\affiliation{
OzGrav-ANU, The Australian Research Council Centre of Excellence for Gravitational Wave Discovery, The Australian National University, Acton, ACT, 2601, Australia
}

\date{\today}

\begin{abstract}
\noindent 
The detection of kilohertz-band gravitational waves promises discoveries in astrophysics, exotic matter, and cosmology. To improve the kilohertz quantum noise--limited sensitivity of interferometric gravitational-wave detectors, we investigate nondegenerate internal squeezing: optical parametric oscillation inside the signal-recycling cavity with distinct signal-mode and idler-mode frequencies.
We use an analytic Hamiltonian model to show that this stable, all-optical technique is tolerant to decoherence from optical detection loss and that it, with its optimal readout scheme, is feasible for broadband sensitivity enhancement.
\end{abstract}
\maketitle

\section{Introduction}

Using the global network of detectors~\cite{abbott2020prospects,AdvancedLIGO:2015,acernese2014advanced,akutsu2018kagra} like the Laser Interferometer Gravitational-Wave Observatory (LIGO)~\cite{AdvancedLIGO:2015} and Virgo~\cite{acernese2014advanced}, much has been learned over the past decade about binary black hole and neutron star mergers from gravitational waves with frequencies around 100~Hz~\cite{cai_2017,Maggiore:2007,GWTC-1:2018,GWTC-2:2020,GWTC-3:2021,vitale2021first}. 
In the future, detecting 1--4~kHz gravitational waves from the coalescence and remnant of binary neutron-star mergers may probe otherwise inaccessible exotic states of matter and further constrain the neutron-star equation-of-state~\cite{PhysRevD.100.104029,miaoDesignGravitationalWaveDetectors2018}. Moreover, \emph{kilohertz} gravitational-wave detection from existing or future detectors~\cite{LIGO_Voyager,NEMO_2020,reitze2019cosmic,maggiore2020science} promises a wealth of discoveries such as determining the origin of low-mass black holes~\cite{PhysRevD.79.044030}, understanding core-collapse supernovae's post-bounce dynamics~\cite{Ott_2009}, and improving non-electromagnetic measurements of the Hubble constant~\cite{PhysRevX.4.041004}.

Quantum shot noise dominates the kilohertz noise for existing gravitational-wave detectors based on the dual-recycled Fabry-Perot Michelson interferometer~\cite{AdvancedLIGO:2015,buikemaSensitivityPerformanceAdvanced2020,PhysRevD.23.1693}. An interferometer's \emph{integrated} quantum noise--limited sensitivity is limited by the circulating optical power and bandwidth of its arm cavities~\cite{mizuno_thesis_1995,miaoFundamentalQuantumLimit2017}. Since increasing the circulating power is technologically challenging~\cite{Brooks_2021,PhysRevLett.114.161102,Barsotti_2018}, improving kilohertz sensitivity requires sacrificing 100~Hz sensitivity unless the above limit can be avoided. Degenerate external squeezing, replacing the vacuum fluctuations entering the readout port with squeezed vacuum, avoids the above limit and reduced the quantum noise by $2.7\pm0.1$~dB at 1.1--1.4~kHz in LIGO~\cite{tseQuantumEnhancedAdvancedLIGO2019,aasietal2013,Ganapathy_2021,PhysRevLett.123.231108,Dooley_2015}. Alone, however, it is not sufficient to achieve the kilohertz sensitivity required for detection~\cite{miaoDesignGravitationalWaveDetectors2018,pageEnhancedDetectionHigh2018}.

To further improve kilohertz sensitivity, two existing proposals are closely related to the present work. Firstly, degenerate internal squeezing (or a ``quantum expander'') uses a nonlinear ``squeezer'' crystal operated degenerately inside the signal-recycling cavity (SRC) of the interferometer to reduce the quantum noise~\cite{korobkoQuantumExpanderGravitationalwave2019,adyaQuantumEnhancedKHz2020}. Secondly, stable optomechanical filtering couples a mechanical mode (e.g.\ a suspended optic) to the optical mode in the signal-recycling cavity; this broadens the arm cavity resonance that limits the kilohertz signal response of the detector (achieving a ``white-light'' cavity)~\cite{liBroadbandSensitivityImprovement2020,liEnhancingInterferometerSensitivity2021,miaoEnhancingBandwidthGravitationalWave2015,WICHT1997431}. These two proposals might each enable kilohertz gravitational-wave detection; the drawbacks are their \emph{high susceptibility to decoherence from optical and mechanical loss, respectively}~\cite{korobkoQuantumExpanderGravitationalwave2019,adyaQuantumEnhancedKHz2020,liBroadbandSensitivityImprovement2020,miaoEnhancingBandwidthGravitationalWave2015}, and requirement for significant technological advances~\cite{ying_2020,pageEnhancedDetectionHigh2018}.

\begin{figure*}[ht!]
    \centering
    \includegraphics[width=\textwidth]{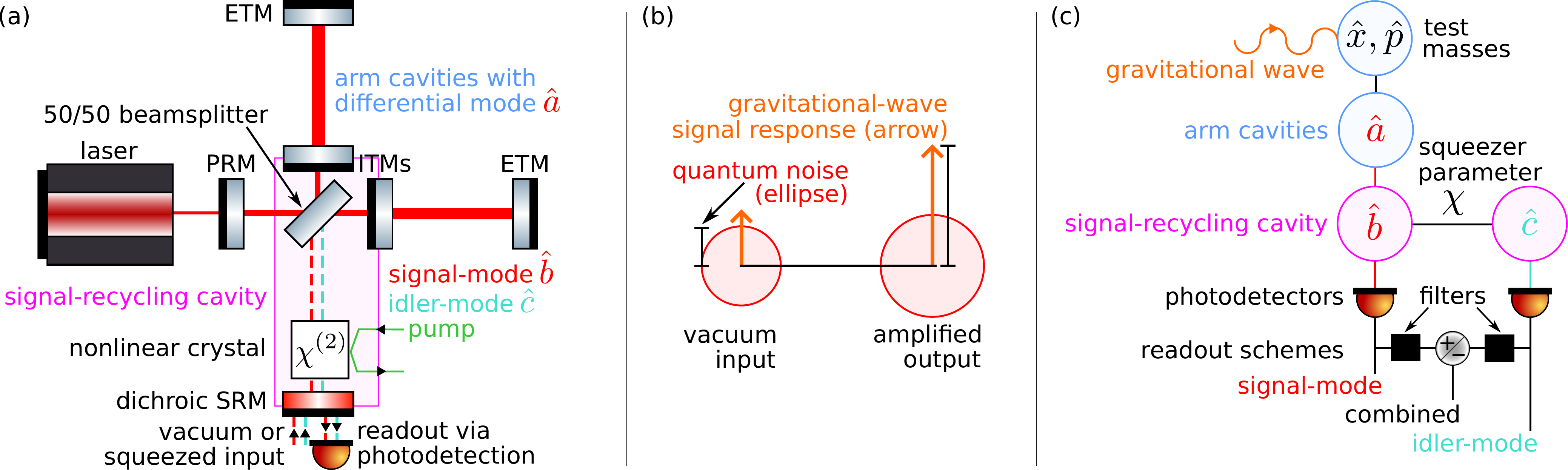}
    \caption{(a) Simplified optical configuration of nondegenerate internal squeezing in a dual-recycled Fabry-Perot Michelson interferometer. Abbreviations are ETM: end test mass, ITM: input test mass, PRM: power-recycling mirror, and SRM: signal-recycling mirror.
    (b) Representation of how signal-mode readout sensitivity is improved by amplifying the signal more than the noise~\cite{danilishinQuantumMeasurementTheory2012}. 
    (c) Mode diagram showing that the system consists of three coupled optical modes ($\hat a, \hat b, \hat c$) and that the gravitational-wave signal indirectly couples into the idler-mode. 
    }
    \label{fig:nIS_config}
\end{figure*}

In this paper, we explore the technique of \emph{nondegenerate internal squeezing} explained below. Although this all-optical technique has an equivalent Hamiltonian to stable optomechanical filtering~\cite{liBroadbandSensitivityImprovement2020,bentley_thesis_2021}, it has not been thoroughly examined to date. We analyse its performance in a future gravitational-wave detector with realistic optical loss and demonstrate further sensitivity improvement using variational readout and optimal filtering~\cite{PhysRevD.65.022002,yap2019generation,PhysRevResearch.3.043079}.

\section{Concept and model}

Nondegenerate internal squeezing consists of a ``squeezer'' crystal with qua\-drat\-ic polarisability ($\chi^{(2)}$) inside the signal-recycling cavity of a dual-recycled Fabry-Perot Michelson interferometer as shown in Fig.~\hyperref[fig:nIS_config]{1(a)}; the squeezer annihilates a pump photon at (angular) frequency $2\omega_0+\Delta$ and creates a pair of photons at ``signal-mode'' (the carrier frequency $\omega_0$) and ``idler-mode'' ($\omega_0+\Delta$ for frequency separation $\Delta\neq0$) frequencies resonant in the signal-recycling cavity. These pairs are Einstein-Podolsky-Rosen (EPR) correlated, \emph{amplified} vacuum states~\cite{schoriNarrowbandFrequencyTunable2002,reidDemonstrationEinsteinPodolskyRosenParadox1989}. 
The signal-mode is coupled to the differential arm mode that contains the gravitational-wave signal~\cite{bond_2010}; the idler-mode frequency is not resonant in the arms.
This technique improves sensitivity by amplifying the gravitational-wave signal more than the quantum noise, as shown in Fig.~\hyperref[fig:nIS_config]{1(b)}, because the signal comes from the arms but the noise comes primarily from the readout port.

\subsection{Analytic model}
\label{sec:model}

We model the system using an established analytic Hamiltonian approach~\cite{danilishinQuantumMeasurementTheory2012, miaoEnhancingBandwidthGravitationalWave2015, liBroadbandSensitivityImprovement2020, korobkoQuantumExpanderGravitationalwave2019,schoriNarrowbandFrequencyTunable2002}. A single-mode ``coupled-cavity'' approximation is valid below the free-spectral range of the arms ($37.5$~kHz~\cite{miaoEnhancingBandwidthGravitationalWave2015}) and gives the differential arm mode (with annihilation Heisenberg operator $\hat a$), signal-mode ($\hat b$), and idler-mode ($\hat c$) shown in Fig.~\hyperref[fig:nIS_config]{1(c)} that evolve according to the Hamiltonian 
\begingroup
\allowdisplaybreaks
\begin{align}\label{eq:nIS_Hamiltonian}
    \hat H &= \hat H_0 + \hat H_\text{int} + \hat H_\text{mech} + \hat H_\text{I/O}\\
    \hat H_0/\hbar &= \omega_0 \hat a^\dag \hat a + \omega_0 \hat b^\dag \hat b+ (\omega_0+\Delta) \hat c^\dag \hat c + (2\omega_0+\Delta) \hat u^\dag \hat u\nonumber\\ 
    \hat H_\text{int} &= i\hbar\omega_s(\hat a\hat b^\dag-\hat a^\dag\hat b) + \hbar g (\hat u \hat b^\dag \hat c^\dag+\hat u^\dag \hat b \hat c)/2\nonumber\\
    \hat H_\text{mech} &= -\alpha (\hat{x}-L_\text{arm}h(t))\left(\hat{a}+\hat{a}^\dag\right)/\sqrt{2}+\hat{p}^2/(2\mu)\nonumber\\
    \frac{\hat H_\text{I/O}}{i\hbar \sqrt{2}} &= \int_{-\infty}^\infty \biggl( \sqrt{\gamma^b_R}\hat{B}^\dag(\omega)\hat{b}
    + \sqrt{\gamma^c_R}\hat{C}^\dag(\omega)\hat{c} + \sqrt{\gamma_a}\hat{N_a}^\dag(\omega)\hat{a}\nonumber\\
    &+ \sqrt{\gamma_b}\hat{N_b}^\dag(\omega)\hat{b} 
    + \sqrt{\gamma_c}\hat{N_c}^\dag(\omega)\hat{c}
    + \text{h.c.} \biggr)\frac{\text{d}\omega}{\sqrt{2\pi}}\,.\nonumber
\end{align}
\endgroup 

Here, $\hat H_0$ describes the uncoupled harmonic behaviour; $\hat H_\text{int}$ describes the optical interaction~\cite{graham1968quantum}; $\hat H_\text{mech}$ describes the gravitational-wave strain ($h(t)$ for time $t$) coupling through the test masses' differential mechanical mode (with free mass position $\hat x$ and momentum $\hat p$) via radiation pressure~\cite{kimble2001conversion}; and $\hat H_\text{I/O}$ describes the readout (intra-cavity loss) rate $\gamma^j_R$ ($\gamma_j$) for $j=b,c$ ($a,b,c$) into vacuum bath modes $\hat B$,$\hat C$ ($\hat{N}_j$) that define the incoming fields $\hat{B}^\text{in}$,$\hat{C}^\text{in}$($\hat{N}^\text{in}_\text{j}$)~\cite{gardiner1985input} where $\gamma = -c/(4L)\log(1-T)$ for $c$ the speed of light, $L$ the cavity length, and $T$ the readout (loss) port transmission. We omit the natural evolution of the vacuum modes for brevity.

In Eq.~\ref{eq:nIS_Hamiltonian}, $\hbar$ is the reduced Plank constant, $\hat u$ is the pump mode, $\omega_s\approx c\sqrt{T_\text{ITM}/(4 L_\text{arm} L_\text{SRC})}$ is the ``sloshing'' frequency~\cite{korobkoQuantumExpanderGravitationalwave2019,thuring2007detuned}, $T_\text{ITM}$ is the input test masses' transmission, $L_\text{arm}$ ($L_\text{SRC}$) is the arm (signal-recycling) cavity length, $\chi^{(2)}$ determines the nonlinear coupling rate $g$~\cite{paschotta1994nonlinear}, $\alpha=\sqrt{2 P_\text{circ} \omega_0 \hbar/(c  L_\text{arm})}$ is the optomechanical coupling rate~\cite{liBroadbandSensitivityImprovement2020}, $P_\text{circ}$ is the circulating (arm) power, and $\mu=M/4$ is the differential mechanical mode's reduced mass (for test mass mass $M$).

For gravitational-wave detectors, the pump power should be kept below the squeezing threshold and a ``reservoir pump'' approximation is valid: $\hat u\mapsto ue^{i\phi}$ where $u$ is the \emph{constant} real amplitude and $\phi$ is the pump phase~\cite{walls_1995,martinelli2001classical,korobkoQuantumExpanderGravitationalwave2019,schoriNarrowbandFrequencyTunable2002}. 
This simplifies the Interaction Frame Heisenberg-Langevin equations-of-motion~\cite{PhysRevA.31.3761,PhysRevA.30.1386} to
\begingroup\allowdisplaybreaks
\begin{align}\label{eq:nIS_EoM}
    \dot{\hat{a}}&=-\omega_s\hat{b}+\frac{i\alpha}{\sqrt{2}\hbar}(\hat{x}-L_\text{arm}h(t)) - \gamma_a \hat{a} + \sqrt{2\gamma_a}\hat{N}_\text{a}^\text{in}\\
    \dot{\hat{b}}&=\omega_s\hat{a} - i\chi e^{i\phi}\hat{c}^\dagger - \gamma^b_\text{tot} \hat{b} + \sqrt{2\gamma^b_R}\hat{B}^\text{in} + \sqrt{2\gamma_b}\hat{N}_\text{b}^\text{in}\nonumber\\
    \dot{\hat{c}}&=-i\chi e^{i\phi}\hat{b}^\dagger - \gamma^c_\text{tot} \hat{c} + \sqrt{2\gamma^c_R}\hat{C}^\text{in} + \sqrt{2\gamma_c}\hat{N}_\text{c}^\text{in}\nonumber\\
    \dot{\hat{x}}&=\frac{1}{\mu}\hat{p},\quad \dot{\hat{p}}=\frac{\alpha}{\sqrt2}(\hat{a}+\hat{a}^\dag)\,.\nonumber
\end{align}
\endgroup
Here, $\chi=gu/2$ is the ``squeezer parameter'' (to be distinguished from the quadratic polarisability, $\chi^{(2)}$), $\gamma^j_\text{tot}=\gamma^j_R+\gamma_j$ for $j=b,c$, and each operator (e.g.\ $\hat a$) is implicitly the fluctuating component (e.g.\ $\delta\hat a(t) = \hat a(t) - \langle\hat a\rangle$ for the time-average $\langle\hat a\rangle$). By solving Eq.~\ref{eq:nIS_EoM} linearly in the Fourier domain of frequency $\Omega$ to find the cavity modes, using input/output relations at the readout port to find the outgoing modes~\cite{PhysRevA.31.3761}, and introducing optical detection loss $R_\text{PD}\in(0,1)$ and vacuum $\hat{N}_\text{j,PD}^\text{in}$ for $j=b,c$, the measured quadratures in the Quadrature Picture (e.g.\ $\hat X_{B_\text{meas}, \theta}=\frac{1}{\sqrt 2}(e^{-i \theta}\hat{B}_\text{meas}+e^{i \theta}\hat{B}_\text{meas}^\dag)$ for the signal-mode), describing the amplitude and phase of the light at the photodetector, are
\begin{align}\label{eq:X_meas}
    \vec{\hat X}_\text{meas}&=\mathbf{T}\vec h + \sum\nolimits_{j\in\{B/C^\text{in}, N_\text{a}^\text{in}, N_\text{b/c}^\text{in}, N_\text{PD}^\text{in}\}} \mathbf{R}_j \vec{\hat X}_j\,.
\end{align}
Here, $\mathbf{T}$ ($\mathbf{R}$) is the resulting signal (noise) transfer matrix describing the relation between an input and the measured output. Each operator-vector contains two quadratures for each of the signal-mode and idler-mode, e.g. 
\begin{equation}
\vec{\hat X}_\text{meas} = (\hat X_{B_\text{meas}, 0}, \hat X_{B_\text{meas}, \frac{\pi}{2}}, \hat X_{C_\text{meas}, 0}, \hat X_{C_\text{meas}, \frac{\pi}{2}})^\text{T},
\end{equation}
but $\vec{h}=\tilde h(\Omega) (1,1,0,0)^\text{T}$ because $\tilde h(\Omega)$ is $h(t)$'s Fourier transform and the idler-mode is not resonant in the arms.

Assuming uncorrelated vacuum noise inputs, the measured quantum noise as a (single-sided) power spectral density matrix is $\mathbf{S}_X(\Omega) = \sum_j \mathbf{R}_j({\mathbf{R}_j}^\ast)^\top$~\cite{danilishinQuantumMeasurementTheory2012}.
By Eq.~\ref{eq:X_meas}, the linear response of the detector to the gravitational wave ($\tilde h(\Omega)$) is $\mathbf{T}(1,1,0,0)^\text{T}$; since its first component is zero, the fixed--readout angle signal-mode readout measures $\hat X_{B_\text{meas},\frac{\pi}{2}}$ with sensitivity~\cite{moore2014gravitational}
\begin{equation}\label{eq:nIS_sigRO_sens}
\sqrt{S_h(\Omega)} = \sqrt{(\mathbf{S}_X(\Omega))_{2,2}}/\abs{\left(\mathbf{T}(1,1,0,0)^\text{T}\right)_2}\,.
\end{equation}

These results reduce to the expected lossless and high arm loss limits~\cite{liBroadbandSensitivityImprovement2020,graham1968quantum}.

\begin{figure} 
    \centering
    \includegraphics[width=\columnwidth]{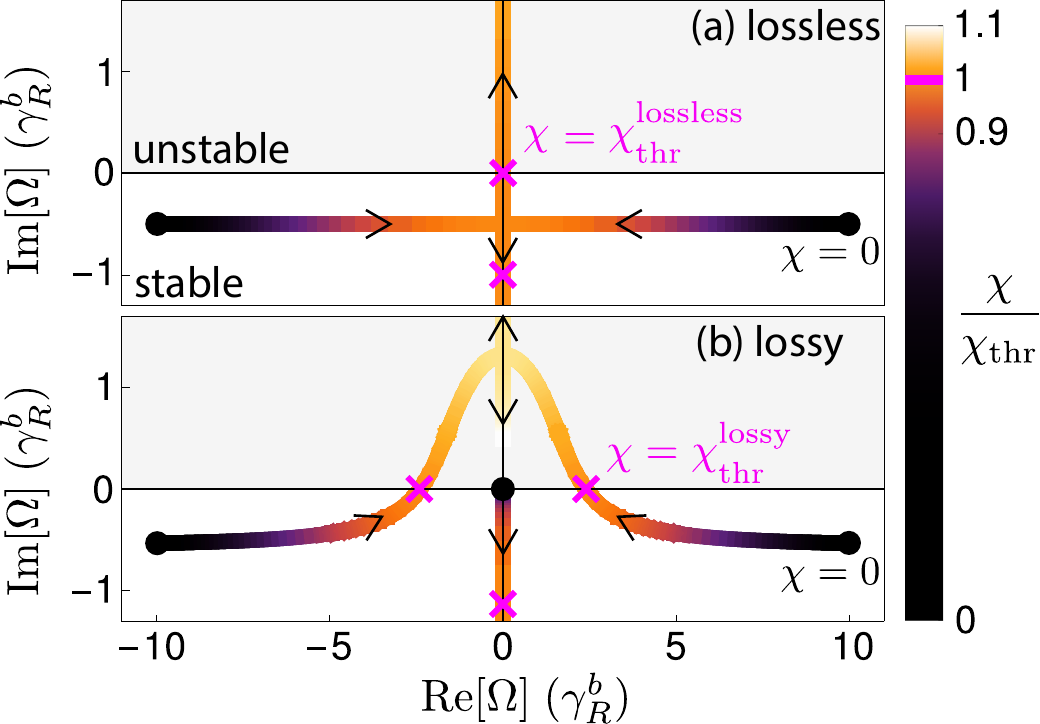}
    \caption{Poles of the transfer functions as the squeezer parameter increases from zero (marked by a dot). Beyond the squeezing threshold (marked by a cross) one or more poles enter the unstable region above the real axis. (a) Lossless and (b) lossy cases using the parameters in Table~\ref{tab:params}. The $\Omega=0$ pole from the free mass approximation is not shown. 
    }
    \label{fig:stability}
\end{figure}

\subsection{Stability and squeezing threshold}

The dynamical stability and squeezing threshold can be determined from the poles of the transfer functions. Here, the transfer functions (e.g.\ the coefficients of $\tilde h(\Omega)$ and $\hat{X}_{B^\text{in}, \frac{\pi}{2}}$ in Eq.~\ref{eq:X_meas}) are rational functions in $\Omega$ with the same denominator for each quadrature and each mode. Moreover, the zeros of the denominator are the same for the signal and noise up to multiplicity and a fixed pole at $\Omega=0$ from the free mass assumption that can be ignored. 
The \emph{system is stable} if all of these poles in $\Omega$ have negative imaginary part~\cite{nise_2019}, which occurs (as shown in Fig.~\ref{fig:stability}) for squeezer parameter below the squeezing threshold given, in the relevant regime $\gamma_a<\gamma^c_\text{tot},\gamma_a\ll\omega_s$, by 
\begin{equation}\label{eq:threshold}
\chi_\text{thr}^\text{lossy}=\sqrt{(\gamma_a+\gamma^b_\text{tot})(\gamma_a+\gamma^c_\text{tot}+\frac{\omega_s^2}{\gamma^b_\text{tot}+\gamma^c_\text{tot}})}\,.
\end{equation}
The system --- in this model --- becomes unstable beyond the squeezing threshold because the reservoir-pump approximation implies unbounded coherent amplification of the cavity modes~\cite{walls_1995,martinelli2001classical}; understanding the system's physical behaviour above threshold would require extending the model beyond this approximation~\cite{xingPumpDepletionParametric2022}. 
This novel method of determining threshold recovers the known values in the lossless ($\chi_\text{thr}^\text{lossless}=\omega_s$~\cite{liBroadbandSensitivityImprovement2020}) and high arm loss ($\sqrt{\gamma^b_\text{tot}\gamma^c_\text{tot}}$~\cite{graham1968quantum}) limits.

\section{Results}

\begin{table}
    \centering
    \begin{tabular}{@{}lll@{}}
    \toprule
    carrier wavelength, $2\pi c/\omega_0$ & \multicolumn{2}{l}{\hspace{1.3cm}2 $\mu\text{m}$} \\
    arm cavity length, $L_\text{arm}$ & \multicolumn{2}{l}{\hspace{1.3cm}4 km} \\
    circulating arm power, $P_\text{circ}$ & \multicolumn{2}{l}{\hspace{1.3cm}3 MW} \\
    test mass mass, $M$ & \multicolumn{2}{l}{\hspace{1.3cm}200 kg} \\
    injected external squeezing & \multicolumn{2}{l}{\hspace{1.3cm}10 dB} \\ 
    intra-cavity loss, $T_{l,a ; b(c)}$ & \multicolumn{2}{l}{\hspace{1.3cm}100 ; 1000 (1000) ppm} \\
    
    detection loss, $R_\text{PD}$ & \multicolumn{2}{l}{\hspace{1.3cm}$10\%$} \\ 
    SRM transmission, $T_{\text{SRM},b(c)}$ & \textbf{0.0152 (0)} & 0.046 (0) \\ 
    SRC length, $L_\text{SRC}$ & \textbf{366.5 m} & 56 m \\ 
    ITM transmission, $T_\text{ITM}$ & \textbf{0.0643} & 0.002 \\ 
    sloshing frequency, $\omega_s$ & \textbf{5 kHz} & 2.256 kHz \\
    readout rate, $\gamma^{b(c)}_R$ & \textbf{0.5 (0) kHz} & 10.038 (0) kHz \\ 
    \bottomrule
    \end{tabular}
    \caption{Baseline parameters for signal-mode readout with deviations shown in boldface next to the corresponding LIGO~Voyager values~\cite{LIGO_Voyager}.
    The idler-mode values are shown in parentheses next to the corresponding signal-mode values and are achievable by means of a dichroic if the frequency separation ($\Delta$) can be made sufficiently large, e.g., using additional cavities. Although 10~dB frequency-dependent external squeezing is injected into the signal-mode (and, later, idler-mode), only ${\sim}7$~dB is measured due to loss.} 
    \label{tab:params}
\end{table}

\begin{figure} 
    \centering
    \includegraphics[width=\columnwidth]{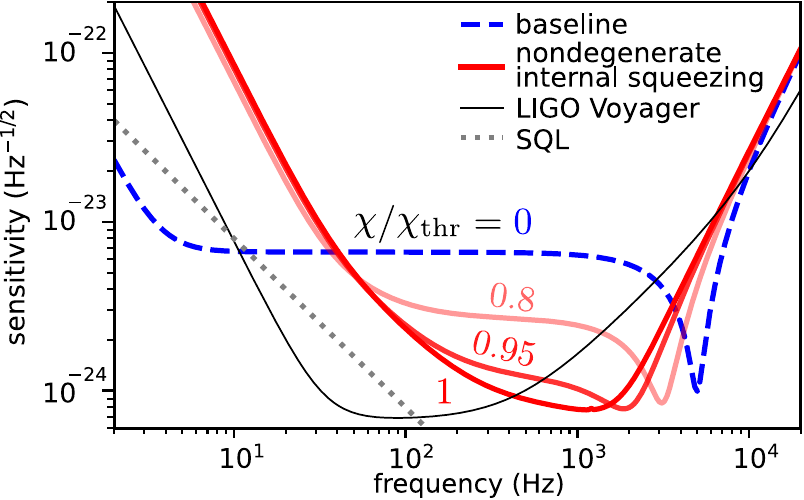} 
    \caption{Sensitivity versus frequency for signal-mode readout with different squeezer parameters. The $\omega_s=5$~kHz feature in the baseline curve is the coupled-cavity pole.
    The baseline curve exceeds the Standard Quantum Limit (SQL)~\cite{danilishinQuantumMeasurementTheory2012} using frequency-dependent external squeezing which is compatible and used with internal squeezing. 
    The parameters in Table~\ref{tab:params} are used. The LIGO~Voyager~\cite{LIGO_Voyager} quantum noise--limited design curve is also shown.} 
    \label{fig:N_S_NSR}
\end{figure}

Nondegenerate internal squeezing improves sensitivity at 40~Hz--4~kHz at the expense of frequencies below 40~Hz as shown in Fig.~\ref{fig:N_S_NSR}. Increasing the squeezer parameter further \emph{improves sensitivity without sacrificing bandwidth or requiring increased circulating power or arm length}. These results use the parameters and realistic optical loss~\cite{zhangBroadbandSignalRecycling2021,Danilishin_2019} in Table~\ref{tab:params} based on LIGO~Voyager~\cite{LIGO_Voyager}; Table~\ref{tab:params} contains a longer signal-recycling cavity than LIGO~Voyager to improve kilohertz sensitivity~\cite{liBroadbandSensitivityImprovement2020}.

Given current estimates of the neutron-star equation-of-state, the sensitivity required to detect a typical binary neutron-star post-merger signal at $50$~Mpc is $\sqrt{S_h}=5\times10^{-25} \mathrm{Hz}^{-1/2}$ from 1--4~kHz~\cite{PhysRevD.100.104029, miaoDesignGravitationalWaveDetectors2018}. With $\chi/\chi_\text{thr}=0.986$ and the parameters in Table~\ref{tab:params} except $T_{\text{SRM},c}=110$~ppm (meaning that technological progress is required), signal-mode readout can achieve this target at ${\sim}1$~kHz. Achieving it across the entire 1--4~kHz band would require reduced loss and increased circulating power, arm length, pump power, and/or injected external squeezing. 

\subsection{Tolerance to decoherence from optical loss}

\begin{figure} 
    \centering
    \includegraphics[width=\columnwidth]{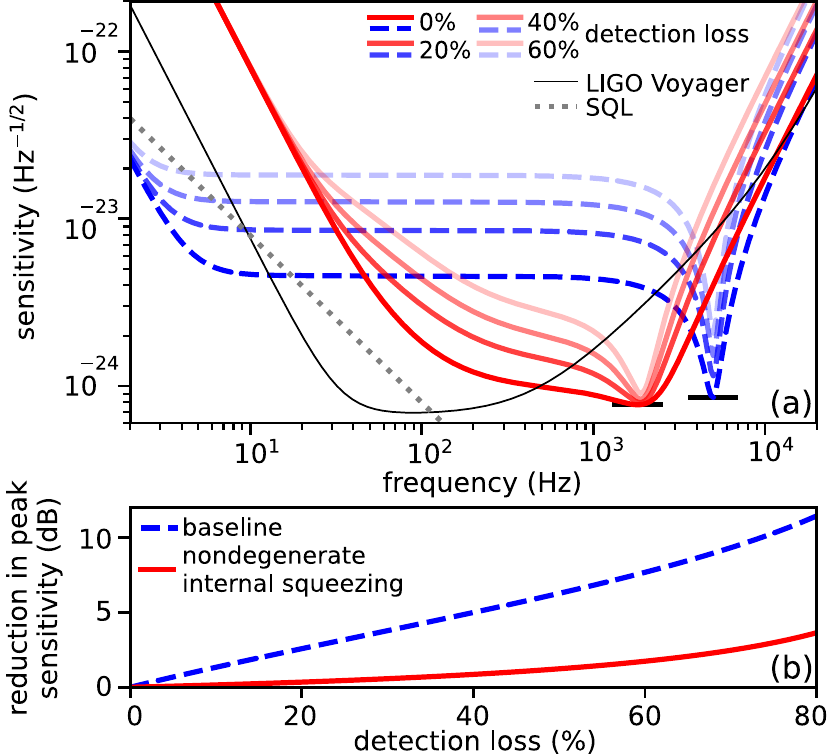} 
    \caption{(a) Sensitivity versus frequency for signal-mode readout with different optical detection loss. (b) Peak sensitivity versus detection loss normalised to the peak sensitivity with $0\%$ detection loss --- higher values indicate greater sensitivity degradation. The parameters, including intra-cavity loss, in Table~\ref{tab:params} are used with $\chi/\chi_\text{thr}=0.95$.} 
    \label{fig:tolerance_to_detection_loss}
\end{figure}

Nondegenerate internal squeezing is more tolerant to decoherence from optical detection loss than a conventional gravitational-wave detector as shown in Fig.~\ref{fig:tolerance_to_detection_loss}. Loss decreases the signal and pulls the quantum noise towards the vacuum level. 
When amplified, however, the signal and noise decrease at approximately the same rate and the sensitivity remains approximately constant. 
In comparison, degenerate internal squeezing experiences worse sensitivity degradation because the squeezed noise increases towards the vacuum level~\cite{korobkoQuantumExpanderGravitationalwave2019,adyaQuantumEnhancedKHz2020,korobkoCompensatingQuantumDecoherenceTalk2021}.

Realistically, signal-mode readout is limited by idler-mode loss which agrees with mechanical idler-mode loss limiting stable optomechanical filtering~\cite{liBroadbandSensitivityImprovement2020,miao2019quantum}. 
To match the sensitivity of stable optomechanical filtering from Ref.~\cite{liBroadbandSensitivityImprovement2020}, which assumes that the mechanical loss parameter-of-interest is a factor of ${\sim}16$ below existing technology, this technique requires a factor of ${\sim}18$ reduction in optical idler-mode loss; the required environmental temperature divided by mechanical quality factor (optical loss) is $6\times 10^{-10}K$~\cite{miaoEnhancingBandwidthGravitationalWave2015} (110~ppm) compared to $9.7\times 10^{-9} K$~\cite{masonetal2019,pageEnhancedDetectionHigh2018} (2000~ppm~\cite{barsottiLIGOdoc2016}) currently possible. This suggests that this technique is a \emph{viable all-optical alternative} to stable optomechanical filtering. This is a key result: this loss-resistant technique is comparable to existing proposals. 

\subsection{Alternative readout schemes}
\label{sec:alternative_readouts}

\begin{figure}
    \centering
    \includegraphics[width=\columnwidth]{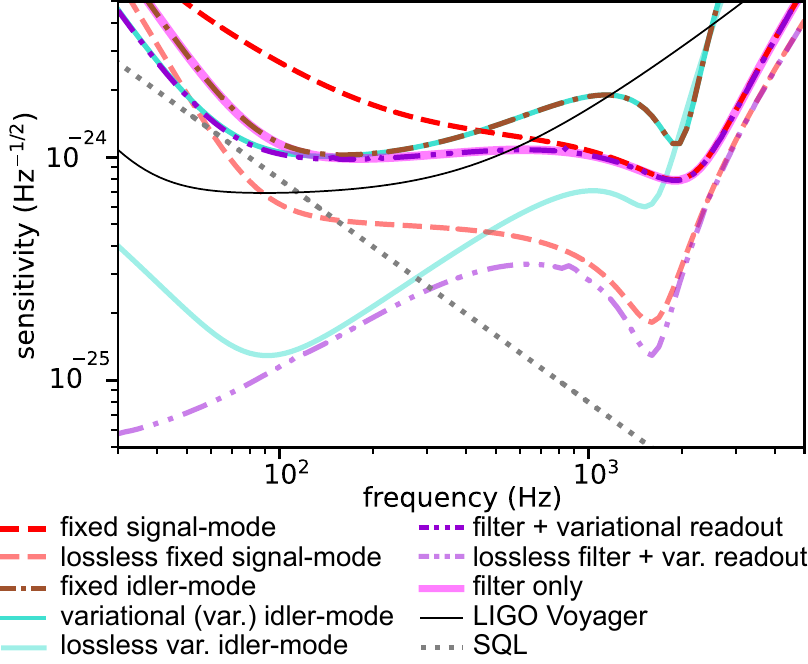}    \caption{Sensitivity versus frequency for alternative readout schemes. The parameters in Table~\ref{tab:params} are used with $\chi/\chi_\text{thr}=0.95$ except $T_{\text{SRM},c}\approx1.54\times10^{-4}$. } 
    \label{fig:alternative_readouts}
\end{figure}

Idler-mode readout is possible, e.g.\ by an additional homodyne readout at the idler-mode frequency to measure $\hat X_{C_\text{meas}, \theta_c}$, because the gravitational-wave signal is coupled in via the squeezer as shown in Fig.~\hyperref[fig:nIS_config]{1(c)}. Since the idler-mode is not directly coupled to the arms, fixed ($\theta_c=\phi$) idler-mode readout improves sensitivity differently than signal-mode readout as shown in Fig.~\ref{fig:alternative_readouts}.
One advantage of idler-mode readout is that the idler-mode wavelength can be chosen to match higher quantum efficiency photodetectors than the $2~\mu\text{m}$ wavelength signal-mode~\cite{singh_2019}.
Realistically, idler-mode readout is limited by signal-mode loss ($T_{l,b}$), however, using idler-mode readout alone with the signal-mode readout port closed does not outperform a signal-mode readout detector. 

Variational readout of each mode, achieved via homodyne readout and a filter cavity~\cite{PhysRevD.65.022002}, can measure $\hat X_{C_\text{vary}} = \hat X_{C_\text{meas}, \theta_c(\Omega)}$ and $\hat X_{B_\text{vary}}$ similar. This improves idler-mode readout, as shown in Fig.~\ref{fig:alternative_readouts}, by reducing the amplified quantum radiation-pressure noise~\cite{danilishinQuantumMeasurementTheory2012,hild2012beyond} using correlations generated ponderomotively at the test masses and coupled from the signal-mode~\cite{PhysRevD.65.022002}. The correlation of the signal-mode quadratures is too low to improve sensitivity for high $\chi/\chi_\text{thr}$ (e.g.\ $0.95$). 

The optimal readout scheme measures the optimal coherent linear combination of the signal-mode and idler-mode, as shown in Fig.~\hyperref[fig:nIS_config]{1(c)}, at each frequency, i.e.\ $\hat X_\text{opt} = \sum_{i\in\{B,C\}} G_i(\Omega)\hat X_{i_\text{vary}}\,,$
where $G_i$ are complex, acausal ``filter'' coefficients, such that $\abs{G_B}^2+\abs{G_C}^2 = 1$, simultaneously numerically optimised with the readout angles ($\theta_b, \theta_c$). This scheme (``filter + variational readout'' in Fig.~\ref{fig:alternative_readouts}) further improves sensitivity via recovering squeezing from the EPR-correlation~\cite{ma_2017,schoriNarrowbandFrequencyTunable2002,liEnhancingInterferometerSensitivity2021}. 
Although decoherence reduces the EPR-correlation, the optimal filter remains more tolerant to detection loss than a conventional detector. In the lossless (i.e.\ no detection or intra-cavity loss) limit, the amplified quantum radiation-pressure noise at 30~Hz can be reduced by up to two orders-of-magnitude as shown in Fig.~\ref{fig:alternative_readouts}. 
Realistically, however, the filter is limited by signal-mode and idler-mode loss, and the optimal filter without variational readout (``filter only'' in Fig.~\ref{fig:alternative_readouts}) achieves the same sensitivity above ${\sim}200$~Hz and is more feasible for a broadband (100~Hz--4~kHz) future gravitational-wave detector.

\section{Conclusions} 

In this paper, we have explored nondegenerate internal squeezing: a viable, all-optical technique to enhance sensitivity. Using an analytic Hamiltonian model, we have found it to (1) be stable, (2) realistically improve sensitivity without sacrificing bandwidth or increasing the circulating power or arm length, and (3) be tolerant to decoherence from optical detection loss --- an advantage over existing proposals.
Using the parameters of a modified LIGO~Voyager, we have shown that optimal filtering without variational readout is this technique's preferred readout scheme out of those considered for kilohertz (1--4~kHz) and broadband (100~Hz--4~kHz) gravitational-wave detection. How to thermally compensate the 100~kW of power incident on the beamsplitter and achieve the large frequency separation required in Table~\ref{tab:params} need further analysis.
This technique may be used in general cavity-based quantum metrology, and our model characterises equivalent Hamiltonian systems, e.g.\ enhanced microwave axion detectors~\cite{liBroadbandSensitivityImprovement2020,MARSH20161,PhysRevX.9.021023}.

\begin{acknowledgments} 

The authors are grateful to the Centre for Gravitational Astrophysics squeezing group for advice during this research and to Xiang~Li for giving access to the results from Ref.~\cite{liBroadbandSensitivityImprovement2020}. Code for this paper was written using Wolfram Mathematica~\cite{mathematica} and Python~\cite{python,ipython,jupyter,numpy,matplotlib} and is openly available at \url{https://github.com/daccordeon/nondegDog}. Fig.~\ref{fig:nIS_config} was illustrated using graphics from Alexander~Franzen~\cite{ComponentLibrary}.
This research was supported by the Australian Research Council under the ARC Centre of Excellence for Gravitational Wave Discovery, Grant No.\ CE170100004. The authors declare no competing interests. This work has been assigned LIGO document number P2200052. 
\end{acknowledgments}

\end{document}